\documentclass[runningheads]{svmult}

\usepackage{graphicx}  
\usepackage{subeqnar}  
\usepackage{multicol}  
\usepackage{physprbb}  

\begin{document}

\title*{Exclusive rare decays of $B$ and $B_c$ mesons 
\protect\newline in a relativistic quark model}

\toctitle{Exclusive rare decays of $B$ and $B_c$ mesons  
in a relativistic quark model}

\titlerunning{Exclusive rare decays of $B$ and $B_c$ mesons  
in a relativistic quark model}

\author{M.A. Ivanov \inst{1} 
\and V.E. Lyubovitskij \inst{2}}

\authorrunning{M.A. Ivanov and V.E. Lyubovitskij }

\institute{Bogoliubov Laboratory of Theoretical Physics, \\  
Joint Institute for Nuclear Research, 141980 Dubna, Russia
\and
Institut f\"ur Theoretische Physik, Universit\"at T\"ubingen, \\ 
Auf der Morgenstelle 14,  D-72076 T\"ubingen, Germany} 

 \maketitle

\begin{abstract}
In these lectures we give, first, the model-independent analysis
of the exclusive rare decays $B\to K \bar l l$ and 
$B_c\to D(D^\ast)\bar l l$ with special emphasis on the cascade decay 
$B_c\to D^\ast(\to D\pi)\bar l l$. 
We derive a four-fold angular decay distribution for this process
in terms of  helicity amplitudes including lepton mass effects.
The  four-fold angular decay distribution allows to define a number 
of physical observables which are amenable to measurement. 
Second, we calculate the relevant form factors within
a relativistic constituent quark model, for the first time 
without employing the impulse approximation. The calculated form
factors are used to evaluate  differential decay rates and
polarization observables. We present results on  
a set of observables  with and without long-distance contributions. 
and compare them with the results of other studies.

\end{abstract}

\section{Introduction}
\label{sec:intr}

The flavor-changing neutral current transitions $B \to K + X$ 
and $B_c\to D(D^\ast) + X$ with $X = \gamma, l^+l^-, \bar\nu\nu$   
are of special interest  because they proceed at the loop  
level in the Standard Model (SM) involving also the top quark.  
They may therefore be used for a determination of the 
Cabibbo-Kobayashi-Maskawa (CKM) matrix elements $V_{tq}$ 
$(q=d,s,b)$. The available experimental measurements of the 
branching ratio of the inclusive radiative $B$-meson decay   
$$\hspace*{-.1cm}
{\rm Br}\left(B\to X_s\gamma\right)=
\left\{
\begin{array}{lr}
\left(3.11\pm 0.80 ({\rm stat}) \pm 0.72 
({\rm syst})\right)\times 10^{-4}  &
\mbox{ALEPH \cite{Barate:1998vz}} \\
 & \\
\left(3.36\pm 0.53 ({\rm stat})\pm 0.42 
({\rm syst})^{+0.50}_{-0.54}
({\rm th})\right)\times 10^{-4}  & 
\mbox{BELLE \cite{Abe:2001hk}} \\
 & \\ 
\left(3.21\pm 0.43 ({\rm stat})\pm 0.27 
({\rm syst})^{+0.18}_{-0.10}
({\rm th})\right)\times 10^{-4}  & 
\mbox{CLEO \cite{Chen:2001fj}} \\
\end{array}
\right.
$$ 
are consistent with the next-to-leading order prediction of the 
standard model (see, e.g.~\cite{Ali:2001jg} and references therein):   
\begin{equation}
{\rm Br}(B\to X_s\gamma)_{\rm SM}=(3.35\pm 0.30)\times 10^{-4}\,. 
\end{equation}  
The decay  $B\to K\,l^+l^-$ $(l=e,\mu)$ has been 
observed by the BELLE Collaboration \cite{Abe:2001dh} 
with a branching ratio of 
\begin{equation}
\label{bkll}
{\rm Br}\left(B\to K\,l^+l^- \right)=
(0.75^{+0.25}_{-0.21}\pm 0.09)\times 10^{-6}\,.
\end{equation}  
The recent observation of the $B_c$ meson by the 
CDF Collaboration at Tevatron in Fermilab \cite{Abe:1998bc} 
raises hopes that one may  also explore the rare decays of 
the bottom-charm meson in the future.  

The theoretical study of the exclusive rare decays proceeds
in two steps. First, the effective Hamiltonian  for such
transitions is derived by calculating the leading and next-to-leading
loop diagrams in the SM and by using the operator product expansion
and renormalization group techniques. The modern status of this part
of the calculation is described  in the review \cite{Buras:1995} 
(and references therein). Second, one needs to evaluate the matrix 
elements of the effective Hamiltonian between  hadronic states. 
This part of the calculation is model dependent since it involves 
nonperturbative QCD. There are many papers  on this subject.
The decay rates, dilepton invariant mass spectra and the 
forward-backforward asymmetry in the decays 
$B\to K\,l^+l^-\,(l=e,\mu,\tau)$ have been investigated in the SM 
and its supersymmetric extensions by using improved form factors 
from light-cone QCD sum rules \cite{Ali:1999mm}. An updated analysis 
of  these decays has been done in \cite{Ali:2001jg} by including  
explicit $O(\alpha_s)$ and $\Lambda_{\rm QCD}/m_b$ corrections. The 
invariant dilepton mass spectrum and the Dalitz plot for the decay 
$B\to K\,l^+l^-$ have been studied in \cite{Greub:1994pi} by using  
quark model form factors. The $B\to K\,l^+l^-$ decay form factors were 
studied via QCD sum rules in \cite{Colangelo:1995jv} and  within the 
lattice-constrained dispersion quark model in \cite{Melikhov:1997wp}. 
Various aspects of these decays were discussed in numerous papers by
Aliev {\it et al.}~\cite{Aliev}. The exclusive semileptonic rare decays 
$B\to K\,l^+l^-$ were analyzed in supersymmetric theories 
in~\cite{Yan:2000dc}. The angular distribution and CP asymmetries in 
the decays $B\to K\pi e^+e^-$ were investigated in~\cite{Kruger:1999xa}. 
The lepton polarization for the inclusive decay $B\to X_s l^+l^-$ was 
discussed in~\cite{Kruger:1996cv} and~\cite{Hewett:1995dk}. The rare 
decays of $B_c\to D(D^\ast)\,l^+l^-$ were studied in~\cite{Geng:2001vy} 
by using the form factors evaluated in the light front and constituent 
quark models.
 
We employ a relativistic quark model \cite{Ivanov:1996pz,Ivanov:1999ic}
to  calculate the decay  form factors. This model is based on an 
effective Lagrangian which describes the coupling of hadrons $H$ to 
their constituent quarks. The coupling strength is determined by the 
compositeness condition $Z_H=0$ \cite{SWH,Efimov:zg} where $Z_H$ is the 
wave function renormalization constant of the hadron $H$. 
One starts with an effective Lagrangian written down in
terms of quark and hadron fields. Then, by using  Feynman
rules, the S-matrix elements describing the hadronic interactions are
given in terms of a set of quark diagrams. In particular, the
compositeness condition enables one to avoid a double counting of
hadronic degrees of freedom. The approach is self-consistent and
universally applicable. All calculations of physical observables 
are straightforward. The model has only a small set of adjustable  
parameters given by  the values of the constituent quark masses and 
the scale parameters that define the size of the distribution of the 
constituent quarks inside a given hadron. The values of the fit 
parameters are within the window of expectations.

The shape of the  vertex functions and the  quark  propagators can in
principle  be found from an analysis of the Bethe-Salpeter and
Dyson-Schwinger equations  as was done e.g. in \cite{Ivanov:1998ms}.
In this paper, however, we choose a  phenomenological approach
where the vertex functions are modelled by a Gaussian form, the size
parameter of which is determined by a fit to the leptonic and radiative
decays of the lowest lying charm and bottom mesons. For the quark 
propagators we use the local representation. In the present calculations 
we do not employ the so-called impulse approximation used 
previously \cite{Ivanov:1999ic}. 
The numerical results obtained with and without the impulse approximation
are close to each other for light-to-light and heavy-to-heavy transitions
but differ considerably from one another for  heavy-to-light transitions 
as e.g. in the  $B\to \pi$ transitions.  

We calculate the form factors of the transition $B\to K$ and use them 
to evaluate differential decay rates and polarization observables with 
and without  long-distance contributions which include the lower-lying 
charmonium states according to \cite{Ali:1991is}. We extend our analysis 
to the exclusive rare decay  $B_c\to D(D^\ast)\bar ll$. We derive a 
four-fold angular decay distribution for the cascade 
$B_c\to D^\ast(\to D \pi)\bar ll$ process in the  helicity frame 
including lepton mass effects following the method outlined 
in \cite{Korner:1989qb}. The four-fold angular decay distribution allows 
one to define  a number of physical observables which are amenable to 
measurement. We compare our results with the ones of other studies. 

\section{Effective Hamiltonian}
\label{sec:ham}

The starting point of the description of the rare exclusive decays is 
the effective Hamiltonian obtained from the SM-diagrams by using the 
operator product expansion and renormalization group techniques. It 
allows one to separate the short-distance contributions and isolate  
them in the Wilson coefficients which can be studied systematically 
within perturbative QCD. The long-distance contributions are contained 
in the matrix elements of local operators. Contrary to the 
short-distance contributions the calculation of such matrix elements 
requires  nonperturbative methods and is therefore model dependent.

We will follow Refs.\cite{Buras:1995} in writing down the analytical 
expressions for the effective Hamiltonian and paper \cite{Ali:1999mm} 
in using the numerical values of the input parameters characterizing 
the short-distance contributions. At the quark level, the rare 
semileptonic decay $b \to s(d) l^+ l^-$ can be described in terms of 
the effective Hamiltonian:  
\begin{equation}
H_{\rm eff} = - \frac{G_F}{\sqrt{2}} \lambda_t   
              \sum_{i=1}^{10} C_i(\mu)  Q_i(\mu) \; . 
\end{equation}
where $\lambda_t \equiv V_{ts(d)}^\dagger V_{tb}$ is the product 
of CKM elements. For example, the standard set \cite{Buras:1995} 
of local operators for $b \to s l^+ l^-$ transition is written as 
\begin{eqnarray} 
\begin{array}{ll} 
Q_1     =  (\bar{s}_{i}  c_{j})_{V-A} \, 
           (\bar{c}_{j}  b_{i})_{V-A} \, ,               &
Q_2     =  (\bar{s} c)_{V-A}  (\bar{c} b)_{V-A}  \, ,              \\
Q_3     =  (\bar{s} b)_{V-A}\sum_q(\bar{q}q)_{V-A} \, ,            &
Q_4     =  (\bar{s}_{i}  b_{j})_{V-A} \sum_q (\bar{q}_{j}
          q_{i})_{V-A} \, ,                                    \\
Q_5     =  (\bar{s} b)_{V-A}\sum_q(\bar{q}q)_{V+A} \, ,            &
Q_6     =  (\bar{s}_{i}  b_{j})_{V-A} 
   \sum_q  (\bar{q}_{j}  q_{i})_{V+A} \, ,               \\
Q_7     =  \frac{e}{8\pi^2} m_b \, \bar{s} \sigma^{\mu\nu}
          (1+\gamma_5) b \, F_{\mu\nu} \, ,                     &
Q_8    =  \frac{g}{8\pi^2} m_b \, \bar{s}_i\sigma^{\mu\nu}
   (1+\gamma_5) {\bf T}_{ij} b_j \,  {\bf G}_{\mu\nu} \, ,          \\
Q_9     = \frac{e}{8\pi^2} (\bar{s} b)_{V-A}  (\bar{l}l)_V  \,     &
Q_{10}  = \frac{e}{8\pi^2} (\bar{s} b)_{V-A}  (\bar{l}l)_A \, 
\end{array}
\end{eqnarray}
where ${\bf G}_{\mu\nu}$ and $F_{\mu\nu}$ are the gluon and photon field 
strengths, respectively; ${\bf T}_{ij}$ are the generators of the 
$SU(3)$ color group; $i$ and $j$ denote color indices (they are omitted 
in the color-singlet currents). Labels $(V\pm A)$ stand for 
$\gamma^\mu(1\pm\gamma^5)$.  $Q_{1,2}$ are current-current operators,
 $Q_{3-6}$ are QCD penguin operators,  $Q_{7,8}$ are "magnetic penguin" 
operators, and $Q_{9,10}$ are semileptonic electroweak penguin operators.

The effective Hamiltonian leads to the free quark $b\to s l^+l^-$-decay 
amplitude:  
\begin{eqnarray}
M(b\to s\ell^+\ell^-) & = & 
\frac{G_F \alpha}{2\sqrt{2}  \pi} \,\lambda_t \, 
\left\{
 C_9^{\rm eff}\,
       \left( \bar{s}  b \right)_{V-A} \,\left(  \bar l l \right)_V
+C_{10}\left( \bar{s}  b \right)_{V-A} \,\left(  \bar l l \right)_A
\right.
\label{free}\\
&-& 
\left. 
\frac{2m_b}{q^2}\,C_7^{\rm eff}\, 
\left( \bar{s}\,i\sigma^{\mu \nu}\,(1+\gamma^5)\, q^\nu \,b \right)\,
                        \left( \bar l l \right)_V
\right\}.\nonumber
\end{eqnarray}
where $C_7^{\rm eff}= C_7 -C_5/3 -C_6$. 
The Wilson coefficient $ C_9^{\rm eff}$ effectively takes 
into account, first, the contributions from the four-quark
operators $Q_i$ (i=1,...,6) and, second, the nonperturbative 
effects coming from the $c\bar c$-resonance contributions
which are as usual parametrized by a Breit-Wigner ansatz 
\cite{Ali:1991is}:

\begin{eqnarray}
C_9^{\rm eff} & = & C_9 + 
C_0 \left\{
h(\hat m_c,  s)+ \frac{3 \pi}{\alpha^2}\,  \kappa\,
         \sum\limits_{V_i = \psi(1s),\psi(2s)}
      \frac{\Gamma(V_i \rightarrow l^+ l^-)\, m_{V_i}}
{  {m_{V_i}}^2 - q^2  - i m_{V_i} \Gamma_{V_i}}\right\} \nonumber\\
&-& \frac{1}{2} h(1,  s) \left( 4 C_3 + 4 C_4 +3 C_5 + C_6\right)\\
&-& \frac{1}{2} h(0,  s) \left( C_3 + 3 C_4 \right) +
\frac{2}{9} \left( 3 C_3 + C_4 + 3 C_5 + C_6 \right)\,.
\nonumber
\end{eqnarray}
where $C_0\equiv 3 C_1 + C_2 + 3 C_3 + C_4+ 3 C_5 + C_6$,
      $\hat m_c=m_c/m_B$, $s=q^2/m_B^2$ and $\kappa=1/C_0$.
Explicit expressions for the function $h(\hat m_c,s)$,
$m_b=m_b(\mu)$ and $\alpha_s(\mu)$ can be found in Refs.~\cite{Buras:1995}.
The numerical values of the input parameters are 
taken from  \cite{Ali:1999mm} and the corresponding values 
of the Wilson coefficients used in the numerical calculations 
are listed in Table~\ref{tab:input}. 

\section{Form factors and differential decay distributions}

We specify our choice of the momenta as $p_1=p_2+k_1+k_2$ with 
$p_1^2=m_1^2$, $p_2^2=m_2^2$ and  $k_1^2=k_2^2=\mu^2$ where $k_1$ and 
$k_2$ are the $l^+$ and $l^-$ momenta, and $m_1$, $m_2$, $\mu$ are 
the masses of initial and final mesons and lepton, respectively.

We define dimensionless form factors by
\begin{eqnarray}
&&
<K(D)(p_2)\,|\,\bar s(d)\, \gamma_\mu\, b\,| B(B_c)(p_1)>= 
F_+(q^2) P_\mu+F_-(q^2) q_\mu\,,
\label{ff}\\
&&\nonumber\\
&&
<K(D)(p_2)\,|\,\bar s(d) \,i\sigma_{\mu\nu}q^\nu\, b\,|\,B(B_c)(p_1)>=
- \frac{1}{m_1+m_2} \, P_\mu^\perp \, q^2\, F_T(q^2)\,,\nonumber\\
&&\nonumber\\
&&
i<D^\ast(p_2,\epsilon_2)\,|\,\bar d\, O_\mu\, b\,|\,B_c(p_1)> =
\frac{1}{m_1+m_2}\,\epsilon_2^{\dagger\nu}\nonumber\\
& &\times \{
-g_{\mu\nu}\,Pq\,A_0(q^2) +P_\mu P_\nu\,A_+(q^2)
+q_\mu P_\nu\,A_-(q^2)+
i\varepsilon_{\mu\nu\alpha\beta} P^\alpha q^\beta 
\,V(q^2)\}\,,
\nonumber\\
&&\nonumber\\
&&i<D^\ast(p_2,\epsilon_2)\,|\,
\bar d\, i\sigma_{\mu\nu}q^\nu(1+\gamma_5)\, b \,|\,B_c(p_1)>= 
\nonumber\\
& & = \epsilon_2^{\dagger\nu}\{ \, g_{\mu\nu}^\perp \,Pq 
\,a_0(q^2) - P_\mu^\perp \, P_\nu\,a_+(q^2)
-i\varepsilon_{\mu\nu\alpha\beta} P^\alpha q^\beta \,g(q^2)\}
\nonumber
\end{eqnarray}
where $P=p_1+p_2$, \, $q=p_1-p_2$, \, 
$P_\mu^\perp \doteq P_\mu - q_\mu Pq/q^2$, \, 
$g_{\mu\nu}^\perp \doteq g_{\mu\nu} - q_\mu q_\nu/q^2$, \,  
and $\epsilon^\dagger_2$ is the 
polarization four-vector of the $D^\ast$.

\begin{table}
\caption{Central values of the input parameters and the corresponding
values of the Wilson coefficients used in the numerical calculations.}  
\begin{center}
\def\arraystretch{1.2}
        \begin{tabular}{|c|c||l|r|}
\hline
 $m_W$                   & 80.41   GeV      & $C_1$    & -0.248   \\   
 $m_Z$                   & 91.1867 GeV      & $C_2$    &  1.107   \\
 $\sin^2 \theta_W $      & 0.2233           & $C_3$    &  0.011   \\
 $m_c$                   & 1.4     GeV      & $C_4$    & -0.026   \\
 $m_t$                   & 173.8   GeV      & $C_5$    &  0.007   \\
 $m_{b,\rm  pole}$       & 4.8     GeV      & $C_6$    & -0.031   \\
 $\mu$                   & $m_{b,\rm pole}$ & $C_7^{\rm eff}$ & -0.313   \\
 $\Lambda_{QCD}$         & 0.220   GeV      & $C_9$    &  4.344   \\
 $\alpha^{-1}$           & 129              & $C_{10}$ & -4.669   \\
 $\alpha_s (m_Z) $       & 0.119            & $C_0$    &  0.362   \\
 $|V^\dagger_{ts} V_{tb}|$  & 0.0385           &          &          \\
 $|V^\dagger_{td} V_{tb}|$  & 0.008            &          &          \\
 $|V^\dagger_{ts} V_{tb}|/|V_{cb}| $ & 1       &          &          \\
\hline
\end{tabular}
\label{tab:input}
\end{center}
\end{table}

The matrix elements of the exclusive transitions 
$B\to K \bar l l$ and $B_c\to D(D^\ast) \bar l l$ 
are written as
\begin{equation}
\label{exclus}
M\left(B(B_c)\to K(D^\ast)\bar ll\right)=
\frac{G_F}{\sqrt{2}}\cdot\frac{\alpha\lambda_t}{2\pi}\,
\left\{
T_1^\mu\,(\bar l\gamma_\mu l)+T_2^\mu\,(\bar l\gamma_\mu\gamma_5 l)
\right\} 
\end{equation}
where the quantities $T_i^\mu$ are expressed through the form
factors and the Wilson coefficients in the following manner:

\noindent
{\bf (a)\,} $B(B_c)\to K(D) \bar ll$-decay:

\begin{eqnarray}
T_i^\mu &=& {\cal F}_+^{(i)}\,P^\mu+ {\cal F}_-^{(i)}\,q^\mu
\hspace{2cm} (i=1,2)\,,
\label{amp_pp}\\
&&\nonumber\\
{\cal F}_+^{(1)} &=& C_9^{\rm eff}\,F_+ + C_7^{\rm eff}\,F_T\, 
\frac{2m_b}{m_1+m_2}\,,
\nonumber\\
{\cal F}_-^{(1)} &=& C_9^{\rm eff}\,F_- - C_7^{\rm eff}\,F_T\,
\frac{2m_b}{m_1+m_2}\,
\frac{Pq}{q^2}\,,
\nonumber\\
&&\nonumber\\
{\cal F}_\pm^{(2)} &=& C_{10}\,F_\pm\,.\nonumber
\end{eqnarray}

\begin{figure}
\begin{center}
\includegraphics[width=11cm]{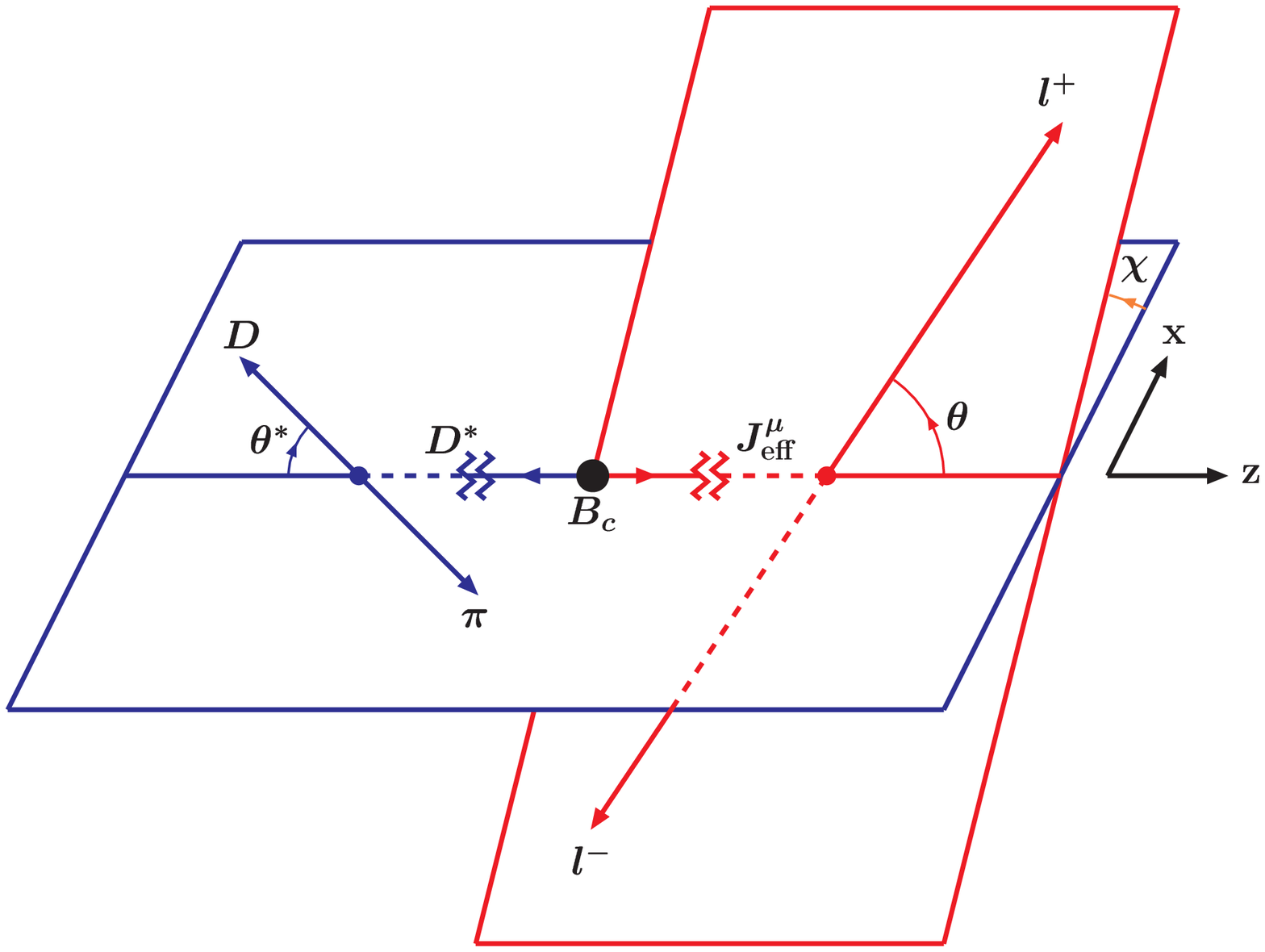}
\end{center}
\caption{Definition of angles $\theta$, $\theta^\ast$ and $\chi$ in  
the cascade decay $B_c\to D^\ast(\to D\pi)\bar l l$.}
\label{fig:bkangl}

\vspace*{6cm}
\includegraphics{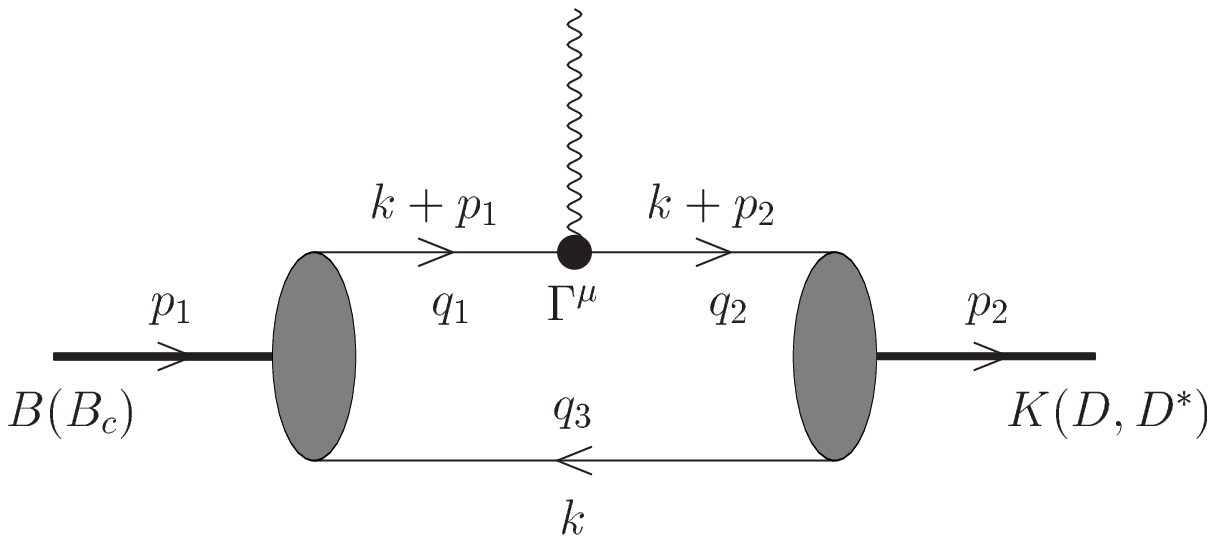}
\caption{Feynman diagram describing the form factors
of the decay  $B(B_c)\to K (D, D^\ast) \bar l l$}
\label{fig:bkformf}
\end{figure}

\noindent
{\bf (b)\,} $B_c\to D^\ast \bar ll$-decay:

\begin{equation}
\label{amp_pv}
T_i^\mu =  T_i^{\mu\nu}\,\epsilon^\dagger_{2\nu}\,, 
\hspace{1cm} (i=1,2)\,,
\end{equation}

\begin{eqnarray*}
 T_i^{\mu\nu} &=& \frac{1}{m_1+m_2}\,
\left\{
- Pq\,g^{\mu\nu}\,A_0^{(i)}+P^\mu P^\nu\,A_+^{(i)}+q^\mu P^\nu\, 
A_-^{(i)}+i\varepsilon^{\mu\nu\alpha\beta}P_\alpha q_\beta\,V^{(i)}  
\right\} 
\\
&&\\
V^{(1)} &=&   C_9^{\rm eff}\,V + C_7^{\rm eff}\,g\, 
\frac{2m_b(m_1+m_2)}{q^2}\,,
\\
&&\\
A_0^{(1)} &=& C_9^{\rm eff}\,A_0 + C_7^{\rm eff}\,a_0\,
\frac{2m_b(m_1+m_2)}{q^2}\,,
\\
&&\\ 
A_+^{(1)} &=& C_9^{\rm eff}\,A_+ + C_7^{\rm eff}\,a_+\,
\frac{2m_b(m_1+m_2)}{q^2}\,,
\\
&&\\
A_-^{(1)} &=& C_9^{\rm eff}\,A_- + C_7^{\rm eff}\,(a_0-a_+)\,
\frac{2m_b(m_1+m_2)}{q^2}\,
\frac{Pq}{q^2}\,,
\\
&&\\
V^{(2)}   &=& C_{10}\,V\,, \hspace{1cm}
A_0^{(2)} = C_{10}\,A_0\,,\hspace{1cm}
A_\pm^{(2)} = C_{10}\,A_\pm\,.
\end{eqnarray*}
Let us first consider the polar angle decay distribution differential
in the momentum transfer squared $q^2$. The polar angle is defined
by the angle between ${\bf q}={\bf p}_1-{\bf p}_2$ and ${\bf k}_1$
($l^+l^-$ rest frame) as shown in Fig.~\ref{fig:bkangl}. One has 

\begin{eqnarray}
\frac{d^2\Gamma}{dq^2 d\cos\theta} &=& 
\frac{|{\bf p_2}|\,v}{(2\pi)^3\,4\,m_1^3}
\cdot\frac{1}{8}\sum\limits_{\rm pol}|M|^2
\label{distr}\\
&&\nonumber\\
&=&
\frac{G^2_F}{(2\pi)^3}\,(\frac{\alpha|\lambda_t|}{2\,\pi})^2
\frac{|{\bf p_2}|\,v } {8 m_1^2}
\cdot\frac{1}{2} 
\biggl\{
 L^{(1)}_{\mu\nu}\cdot (H^{\mu\nu}_{11}+H^{\mu\nu}_{22})
\nonumber\\
&&\nonumber\\
&-&-\frac{1}{2}\,L^{(2)}_{\mu\nu}\cdot 
(q^2\,H^{\mu\nu}_{11}+  (q^2-4\mu^2)\,H^{\mu\nu}_{22})
+L^{(3)}_{\mu\nu}\cdot (H^{\mu\nu}_{12}+H^{\mu\nu}_{21})
\biggr\}
\nonumber
\end{eqnarray}
where  $|{\bf p_2}|=\lambda^{1/2}(m_1^2,m_2^2,q^2)/2m_1$ is
the momentum of the final meson and $v=\sqrt{1-4\mu^2/q^2}$ 
is the lepton velocity both given in the $B(B_c)$-rest frame. 
We have introduced lepton and hadron tensors as

\begin{eqnarray}
L^{(1)}_{\mu\nu} &=& k_{1\mu} k_{2\nu}+ k_{2\mu} k_{1\nu}\,,
\hspace{1cm}
L^{(2)}_{\mu\nu}= g_{\mu\nu}\,,
\hspace{1cm}
L^{(3)}_{\mu\nu}= i\varepsilon_{\mu\nu\alpha\beta}k_1^\alpha k_2^\beta\,,
\nonumber\\
&&\label{tensors}\\
H_{ij}^{\mu\nu} & = &T_i^\mu\,T_j^{\dagger\nu}\,.
\nonumber
\end{eqnarray}

\section{Helicity amplitudes and two-fold distributions}

The Lorentz contractions in Eq.~(\ref{distr}) can be  evaluated in terms of
helicity amplitudes as described  in \cite{Korner:1989qb}.
First, we define an orthonormal and complete helicity basis
$\epsilon^\mu(m)$ with the three spin 1 components  orthogonal to
the momentum transfer $q^\mu$, i.e. $\epsilon^\mu(m) q_\mu=0$ for $m=\pm,0$,
and the spin 0 (time)-component $m=t$ with
$\epsilon^\mu(t)= q^\mu/\sqrt{q^2}$.   

The orthonormality and completeness properties read
 
\begin{equation}
\label{orthonorm}
\epsilon^\dagger_\mu(m)\epsilon^\mu(n)=g_{mn}\,, \hspace{1cm}
\epsilon_\mu(m)\epsilon^{\dagger}_{\nu}(n)g_{mn}=g_{\mu\nu}
\end{equation}
with $(m,n=t,\pm,0)$ and $g_{mn}={\rm diag}\,(\,+\,,\,\,-\,,\,\,-\,,\,\,-\,)$.
We include the time component polarization vector $\epsilon^\mu(t)$
in the set because we want to discuss lepton mass effects in the following.

Using the completeness property we rewrite the contraction
of the lepton and hadron tensors in Eq.~(\ref{distr}) according to

\begin{eqnarray}
L^{(k)\mu\nu}H_{\mu\nu}^{ij} &=& 
L_{\mu'\nu'}^{(k)}g^{\mu'\mu}g^{\nu'\nu}H_{\mu\nu}^{ij}
= L_{\mu'\nu'}^{(k)}\epsilon^{\mu'}(m)\epsilon^{\dagger\mu}(m')g_{mm'}
\epsilon^{\dagger \nu'}(n)\epsilon^{\nu}(n')g_{nn'}H_{\mu\nu}^{ij}
\nonumber\\
&&\nonumber\\
&=& L^{(k)}(m,n)g_{mm'}g_{nn'}H^{ij}(m',n')
\label{contraction}
\end{eqnarray}
where we have introduced the lepton and hadron tensors in the space
of the helicity components

\begin{eqnarray}
\label{hel_tensors}
L^{(k)}(m,n) &=& \epsilon^\mu(m)\epsilon^{\dagger \nu}(n)
L^{(k)}_{\mu\nu}, \hspace{1cm}
H^{ij}(m,n) = \epsilon^{\dagger \mu}(m)\epsilon^\nu(n)H^{ij}_{\mu\nu}.
\end{eqnarray}
The point is that the two tensors can be evaluated in two different
Lorentz systems. The lepton tensors $L^{(k)}(m,n)$ will be evaluated
in the $\bar ll$-CM system whereas the hadron tensors $H^{ij}(m,n)$
will be evaluated in the $B(B_c)$ rest system.

In the $B(B_c)$ rest frame one has 

\begin{eqnarray}
p^\mu_1 &=& (\,m_1\,,\,\,0,\,\,0,\,\,0\,)\,,
\nonumber\\
p^\mu_2 &=& (\,E_2\,,\,\,0\,,\,\,0\,,\,\,-|{\bf p_2}|\,)\,,
\\
q^\mu   &=& (\,q_0\,,\,\,0\,,\,\,0\,,\,\,|{\bf p_2}|\,)\,,
\nonumber
\end{eqnarray}
where $E_2 = (m_1^2+m_2^2-q^2)/2 m_1$ and $q_0=(m_1^2-m_2^2+q^2)/2 m_1$.
In the $B(B_c)$-rest frame the polarization vectors of the effective 
current  read

\begin{eqnarray}
\epsilon^\mu(t)&=&
\frac{1}{\sqrt{q^2}}(\,q_0\,,\,\,0\,,\,\,0\,,\,\,|{\bf p_2}|\,)
\,,\nonumber\\
\epsilon^\mu(\pm) &=& 
\frac{1}{\sqrt{2}}(\,0\,,\,\,\mp 1\,,\,\,-i\,,\,\,0\,)\,,
\label{hel_basis}\\
\epsilon^\mu(0) &=&
\frac{1}{\sqrt{q^2}}(\,|{\bf p_2}|\,,\,\,0\,,\,\,0\,,\,\,q_0\,)\,.
\nonumber
\end{eqnarray}
Using this basis one can express the components of the hadronic
tensors through the invariant form factors defined in  Eq.~(\ref{ff}). 

\vspace{0.5cm}
\noindent
(a) $B(B_c)\to K(D)$ transition:

\begin{equation}
\label{hel_pp}
H^{ij}(m,n)= \left(\epsilon^{\dagger \mu}(m)T^i_\mu\right)\cdot 
        \left(\epsilon^{\dagger \nu}(n)T^j_\nu\right)^\dagger\equiv
H^i(m)H^{\dagger j}(n)
\end{equation}
The helicity form factors $H^i(m)$ are given in terms of the
invariant form factors. One has

\begin{eqnarray}
H^i(t) &=& \frac{1}{\sqrt{q^2}}(Pq\, {\cal F}^i_+ + q^2\, {\cal F}^i_-)\,,
\nonumber\\
H^i(\pm) &=& 0\,,
\label{hel_pp1}\\
H^i(0) &=& \frac{2\,m_1\,|{\bf p_2}|}{\sqrt{q^2}} \,{\cal F}^i_+ \,.
\nonumber
\end{eqnarray} 

\vspace{0.5cm}
\noindent
(b) $B_c\to D^\ast$ transition:

\begin{eqnarray} 
H^{ij}(m,n) &=&  
\epsilon^{\dagger \mu}(m) \epsilon^{ \nu}(n)H^{ij}_{\mu\nu}
=
\epsilon^{\dagger \mu}(m) \epsilon^{ \nu}(n) 
T^i_{\mu\alpha} 
\left(-g^{\alpha\beta}+\frac{p_2^\alpha p_2^\beta}{m_2^2}\right)
T^{\dagger j}_{\beta\nu}
\nonumber\\
&=&
\epsilon^{\dagger \mu}(m) \epsilon^{ \nu}(n) 
T^i_{\mu\alpha}\epsilon_2^{\dagger\alpha}(r)
\epsilon_2^{\beta}(s)\delta_{rs}
T^{\dagger j}_{\beta\nu}
\label{hel_vv}\\
&=&
\epsilon^{\dagger \mu}(m)\epsilon_2^{\dagger\alpha}(r)
T^i_{\mu\alpha} \cdot
\left(\epsilon^{\dagger \nu}(n)\epsilon_2^{\dagger\beta}(s)T^j_{\nu\beta}
\right)^\dagger\delta_{rs}
=H^i(m)H^{\dagger \,j}(n).
\nonumber
\end{eqnarray} 
From angular momentum conservation one has 
$r=m$ and $s=n$ for $m,n=\pm,0$ and $r,s=0$ for $m,n=t$.
For further evaluation one needs to specify the helicity components
$\epsilon_2(m)$ $(m=\pm,0)$ of the polarization vector of the $D^\ast$.
They read
\begin{equation}
\label{vect_pol}
\epsilon^\mu_2(\pm) = 
\frac{1}{\sqrt{2}}(0\,,\,\,\pm 1\,,\,\,-i\,,\,\,0\,)\,,
\hspace{1cm}
\epsilon^\mu_2(0) = 
\frac{1}{m_2}(|{\bf p_2}|\,,\,\,0\,,\,\,0\,,\,\,-E_2\,)\,.
\end{equation}
They satisfy the  orthonormality and completeness properties:
\begin{equation}
\epsilon_2^{\dagger\mu}(r)\epsilon_{2\mu}(s)=-\delta_{rs},
\hspace{1cm}
\epsilon_{2\mu}(r)\epsilon^\dagger_{2\nu}(s)\delta_{rs}=
-g_{\mu\nu}+\frac{p_{2\mu}p_{2\nu}}{m_2^2}.
\end{equation}

Finally one obtains the non-zero components of the hadron tensors 
\begin{eqnarray}
H^i(t) &=& 
\epsilon^{\dagger \mu}(t)\epsilon_2^{\dagger \alpha}(0)T^i_{\mu\alpha}
\,=\,
\frac{1}{m_1+m_2}\frac{m_1\,|{\bf p_2}|}{m_2\sqrt{q^2}}
\left(Pq\,(-A^i_0+A^i_+)+q^2 A^i_-\right),
\nonumber\\
H^i(\pm) &=& 
\epsilon^{\dagger \mu}(\pm)\epsilon_2^{\dagger \alpha}(\pm)T^i_{\mu\alpha}
\,=\,
\frac{1}{m_1+m_2}\left(-Pq\, A^i_0\pm 2\,m_1\,|{\bf p_2}|\, V^i \right),
\label{hel_vv1}\\
H^i(0) &=&  
\epsilon^{\dagger \mu}(0)\epsilon_2^{\dagger \alpha}(0)T^i_{\mu\alpha}
\nonumber\\
&=&
\frac{1}{m_1+m_2}\frac{1}{2\,m_2\sqrt{q^2}} 
\left(-Pq\,(m_1^2+m_2^2-q^2)\, A^i_0
+4\,m_1^2\,|{\bf p_2}|^2\, A^i_+\right).\nonumber
\end{eqnarray} 

The lepton tensors $L^{(k)}(m,n)$ are evaluated 
in the $\bar ll$-CM system ${\bf k}_1+{\bf k}_2=0$.
One has (see Fig.~\ref{fig:bkangl})

\begin{eqnarray}
q^\mu   &=& (\,\sqrt{q^2}\,,\,\,0\,,\,\,0\,,\,\,0\,)\,,
\nonumber\\
k^\mu_1 &=& 
(\,E_1\,,\,\, |{\bf k_1}|\sin\theta\cos\chi\,,\,\, 
|{\bf k_1}| \sin\theta\sin\chi\,,\,\,|{\bf k_1}| \cos\theta\,)\,,
\label{kpi_basis}\\
k^\mu_2 &=& (\,E_1\,,\,\,-|{\bf k_1}|\sin\theta\cos\chi\,,\,\,
-|{\bf k_1}|\sin\theta\sin\chi\,,\,\,-|{\bf k_1}|\cos\theta\,)\,,
\nonumber
\end{eqnarray}
with $E_1=\sqrt{q^2}/2$ and $|{\bf k_1}|=\sqrt{q^2-4\mu^2}/2$.
The longitudinal and time component polarization vectors 
in the $\bar l l$ rest frame 
can be read off from Eq.~(\ref{hel_basis}) and are given by
$\epsilon^\mu(0)=(0,0,0,1)$ and $\epsilon(t)=(1,0,0,0)$ whereas the
transverse parts remain unchanged from Eq.~(\ref{hel_basis}).

The differential $(q^2,\cos\theta)$ distribution finally reads 
\begin{eqnarray}
\frac{d\Gamma(H_{in}\to H_f\bar l l)}{dq^2d(\cos\theta)} &=&\,
\frac{3}{8}\,(1+\cos^2\theta)\cdot
\frac{1}{2}\left(\frac{d\Gamma_U^{11}}{d q^2}+
                   \frac{d\Gamma_U^{22}}{d q^2}\right)
\label{distr2}\\
&&\nonumber\\
&+&\frac{3}{4}\,\sin^2\theta\cdot\frac{1}{2}
\left(\frac{d\Gamma_L^{11}}{d q^2}+
                   \frac{d\Gamma_L^{22}}{d q^2}\right)\nonumber\\
&&\nonumber\\
&-&\,v \cdot\frac{3}{4}\cos\theta\cdot\frac{d\Gamma_P^{12}}{d q^2}
\nonumber\\
&&\nonumber\\
&+&\frac{3}{4}\,\sin^2\theta\cdot\frac{1}{2} 
\frac{d\tilde\Gamma_U^{11}}{d q^2} - \frac{3}{8}\,
(1+\cos^2\theta)\cdot\frac{d\tilde\Gamma_U^{22}}{d q^2}\nonumber\\
&&\nonumber\\
&+&\frac{3}{2}\,\cos^2\theta\cdot\frac{1}{2}
\frac{d\tilde\Gamma_L^{11}}{d q^2}
-\frac{3}{4}\,\sin^2\theta\cdot\frac{d\tilde\Gamma_L^{22}}{d q^2}
+\frac{1}{4}\,\frac{d\tilde\Gamma_S^{22}}{d q^2}\,.\nonumber
\end{eqnarray}

Integrating over $\cos\theta$ one obtains 
\begin{eqnarray}
\frac{d\Gamma(H_{in} \to H_f \bar l l)}{dq^2} &=&\,
\frac{1}{2}
\left(
 \frac{d\Gamma_U^{11}}{d q^2}+\frac{d\Gamma_U^{22}}{d q^2}
+\frac{d\Gamma_L^{11}}{d q^2}+\frac{d\Gamma_L^{22}}{d q^2}
\right)
\label{distr1}\\
&&\nonumber\\
&+&\frac{1}{2}\frac{d\tilde\Gamma_U^{11}}{d q^2} 
-\frac{d\tilde\Gamma_U^{22}}{d q^2}
+\frac{1}{2}\frac{d\tilde\Gamma_L^{11}}{d q^2}
-\frac{d\tilde\Gamma_L^{22}}{d q^2}
+\frac{1}{2}\, \frac{d\tilde\Gamma_S^{22}}{d q^2}\,,
\nonumber
\end{eqnarray}
where the partial helicity rates $d\Gamma_X^{ij}/dq^2$ and 
                           $d\tilde\Gamma_X^{ij}/dq^2$
($X=U,L,P,S;\,i,j=1,2$) are defined as 

\begin{equation}
\label{hel_rate1}
\frac{d\Gamma_X^{ij}}{dq^2} = \frac{G^2_F}{(2\pi)^3} 
\left(\frac{\alpha|\lambda_t|}{2\pi}\right)^2
\frac{|{\bf p_2}|\,q^2\,v}{12\,m_1^2}\,H_X^{ij} \,,
\hspace{1cm}
\frac{d\tilde\Gamma_X^{ij}}{dq^2} =
\frac{2\,\mu^2}{q^2}\,\frac{d\Gamma_X^{ij}}{dq^2} \,.
\end{equation}
The relevant bilinear combinations of the helicity amplitudes
are defined in Table~\ref{tab:helicity}.

\begin{table}
\begin{center}
\caption{Bilinear combinations of the helicity amplitudes that 
enter in the four-fold decay distribution Eq.~(\ref{hel_rate1}).} 
\vspace*{.5cm}
\def\arraystretch{1.8}
\begin{tabular}{|l|l|l|}
\hline
Definition & Property & Title \\
\hline
$ H_{U}^{ij}={\rm Re}\left(H_{+}^i H_{+}^{\dagger\,j}\right)
            +{\rm Re}\left(H_{-}^i H_{-}^{\dagger\,j}\right) $ &
$ H_{U}^{ij}=  H_{U}^{ji}$ & {\bf U}npolarized-transverse \\
$ H_{IU}^{ij}={\rm Im}\left(H_{+}^i H_{+}^{\dagger\,j}\right)
             +{\rm Im}\left(H_{-}^i H_{-}^{\dagger\,j}\right) $ &
$ H_{IU}^{ij}=-H_{IU}^{ji}$ &  \\
\hline
$ H_{P}^{ij}={\rm Re}\left(H_{+}^i H_{+}^{\dagger\,j}\right)
            -{\rm Re}\left(H_{-}^i H_{-}^{\dagger\,j}\right) $ &
$ H_{P}^{ij}=  H_{P}^{ji}$ & {\bf P}arity-odd \\
$ H_{IP}^{ij}={\rm Im}\left(H_{+}^i H_{+}^{\dagger\,j}\right)
             -{\rm Im}\left(H_{-}^i H_{-}^{\dagger\,j}\right) $ &
$ H_{IP}^{ij}=-H_{IP}^{ji}$ &  \\
\hline
$ H_{T}^{ij}={\rm Re}\left(H_{+}^i H_{-}^{\dagger\,j}\right)$ & & 
{\bf T}ransverse-interference \\
$ H_{IT}^{ij}={\rm Im}\left(H_{+}^i H_{-}^{\dagger\,j}\right)$ & & \\
\hline
$ H_{L}^{ij}={\rm Re}\left(H_{0}^i H_{0}^{\dagger\,j}\right) $ &
$ H_{L}^{ij}= H_{L}^{ji}$ & {\bf L}ongitudinal \\
$ H_{IL}^{ij}={\rm Im}\left(H_{0}^i H_{0}^{\dagger\,j}\right)$ &
$ H_{IL}^{ij}= -H_{IL}^{ji}$ & \\
\hline
$ H_{S}^{ij}=3\,{\rm Re}\left(H_{t}^i H_{t}^{\dagger\,j}\right) $ &
$ H_{S}^{ij}= H_{S}^{ji}$ & {\bf S}calar \\
$ H_{IS}^{ij}=3\,{\rm Im}\left(H_{t}^i H_{t}^{\dagger\,j}\right)$ &
$ H_{IS}^{ij}= -H_{IS}^{ji}$ & \\
\hline
$ H_{SL}^{ij}={\rm Re}\left(H_{t}^i H_{0}^{\dagger\,j}\right) $ & & 
{\bf S}calar-{\bf L}ongitudinal-\\[-2.5mm]
& & interference \\[-1.5mm]
$ H_{ISL}^{ij}={\rm Im}\left(H_{t}^i H_{0}^{\dagger\,j}\right)$ & & \\
\hline
$ H_{I}^{ij}=\frac{1}{2}\,
 \left[{\rm Re}\left(H_{+}^i H_{0}^{\dagger\,j}\right)
      +{\rm Re}\left(H_{-}^i H_{0}^{\dagger\,j}\right)\right]$ & &
  transverse-longitudinal-\\[-2.5mm]
& & {\bf I}nterference  \\[-1.5mm]
$ H_{II}^{ij}=\frac{1}{2}\,
\left[{\rm Im}\left(H_{+}^i H_{0}^{\dagger\,j}\right)
     +{\rm Im}\left(H_{-}^i H_{0}^{\dagger\,j}\right)\right]$ 
&  &  \\
\hline
$ H_{A}^{ij}=\frac{1}{2}\,
\left[{\rm Re}\left(H_{+}^i H_{0}^{\dagger\,j}\right)
     -{\rm Re}\left(H_{-}^i H_{0}^{\dagger\,j}\right)\right]$ 
&  & parity-{\bf A}symmetric  \\
$ H_{IA}^{ij}=\frac{1}{2}\,
\left[{\rm Im}\left(H_{+}^i H_{0}^{\dagger\,j}\right)
     -{\rm Im}\left(H_{-}^i H_{0}^{\dagger\,j}\right)\right]$ 
&  &  \\
\hline
$ H_{ST}^{ij}=\frac{1}{2}\,
\left[{\rm Re}\left(H_{+}^i H_{t}^{\dagger\,j}\right)
     +{\rm Re}\left(H_{-}^i H_{t}^{\dagger\,j}\right)\right]$ 
& & {\bf S}calar-{\bf T}ransverse-\\[-2.5mm]
& & interference  \\[-1.5mm]
$ H_{IST}^{ij}=\frac{1}{2}\,
\left[{\rm Im}\left(H_{+}^i H_{t}^{\dagger\,j}\right)
     +{\rm Im}\left(H_{-}^i H_{t}^{\dagger\,j}\right)\right]$ 
&  &  \\
\hline
$ H_{SA}^{ij}=\frac{1}{2}\,
\left[{\rm Re}\left(H_{+}^i H_{t}^{\dagger\,j}\right)
     -{\rm Re}\left(H_{-}^i H_{t}^{\dagger\,j}\right)\right]$ 
& & {\bf S}calar-{\bf A}symmetric-\\[-2.5mm]
& & interference  \\[-1.5mm]
$ H_{ISA}^{ij}=\frac{1}{2}\,
\left[{\rm Im}\left(H_{+}^i H_{t}^{\dagger\,j}\right)
     -{\rm Im}\left(H_{-}^i H_{t}^{\dagger\,j}\right)\right]$ 
&  &  \\
\hline
\end{tabular}
\label{tab:helicity}
\end{center}
\end{table}

\section{The four-fold angle distribution in the cascade decay \\
$B_c\to D^\ast(\to D\pi) \bar ll$.}

The lepton-hadron correlation function $L_{\mu\nu}H^{\mu\nu}$ reveals 
even more structure when one uses the cascade decay 
$B_c\to D^\ast(\to D\pi)\bar l l$ to analyze the polarization of the 
$D^\ast$. The hadron tensor now reads
\begin{equation}
H^{ij}_{\mu\nu}=T^i_{\mu\alpha}(T^j_{\nu\beta})^\dagger
\frac{3}{2\,|{\bf p_3}|}{\rm Br}(K^\ast\to K\pi)p_{3\alpha'}p_{3\beta'}
S^{\alpha\alpha'}(p_2)S^{\beta\beta'}(p_2)
\end{equation}
where 
$S^{\alpha\alpha'}(p_2)=-g^{\alpha\alpha'}
+p_2^\alpha p_2^{\alpha'}/m_2^2$ 
is the standard spin 1 tensor,
$p_2=p_3+p_4$, $p_3^2=m_D^2$, $p_4^2=m_\pi^2$, and  $p_3$ and $p_4$
are the momenta of the $D$ and the $\pi$, respectively. 
The relative configuration of the ($D,\pi$)- and ($\bar l l$)-planes
is shown in Fig.~\ref{fig:bkangl}. 

In the rest frame of the $D^\ast$ one has 
\begin{eqnarray}\label{casc_mom}
p^\mu_2 &=& (m_{D^\ast},\vec{0}), \\
p^\mu_3 &=& (\,E_D\,,\,\,|{\bf p_3}|\,\sin\theta^\ast\,,\,\,0\,,\,\,
                    -|{\bf p_3}|\,\cos\theta^\ast\,)\,,\nonumber\\   
p^\mu_4 &=& (\,E_\pi\,,\,\,-|{\bf p_3}|\,\sin\theta^\ast\,,\,\,0\,,\,\,
                        |{\bf p_3}|\,\cos\theta^\ast\,)\, ,\nonumber\\ 
|{\bf p_3}| &=& 
\lambda^{1/2}(m_{D^\ast}^2,m_D^2,m_\pi^2)/(2\,m_{D^\ast})\, . \nonumber
\end{eqnarray}
Without loss of generality we set the azimuthal angle $\chi^\ast$
of the $(D,\pi)$-plane to zero. 
According to Eq.~(\ref{vect_pol}) the rest frame polarization vectors
of the $D^\ast$ are given by
\begin{equation}
\label{casc_pol}
\epsilon^\mu_2(\pm) = \frac{1}{\sqrt{2}}(\,0\,,\,\,\pm\,,\,\,
-i\,,\,\,0\,)\,,\hspace{1cm}
\epsilon^\mu_2(0) = (\,0\,,\,\,0\,,\,\,0\,,\,\,-1)\,.
\end{equation}
The spin 1 tensor $S^{\alpha\alpha'}(p_2)$ is then written as 
\begin{equation}
S^{\alpha\alpha'}(p_2)=-g^{\alpha\alpha'}
+\frac{p_2^\alpha p_2^{\alpha'}}{m_2^2}
=\sum\limits_{m=\pm,0}\epsilon_2^\alpha(m)\epsilon_2^{\dagger\alpha'}(m)
\end{equation}
Following basically the same trick as in Eq.~(\ref{contraction})
the contraction of the lepton and hadron tensors may be
written through helicity components as
\begin{eqnarray} 
L^{(k)\mu\nu}H^{ij}_{\mu\nu}&=&
\epsilon^{\mu'}(m)\epsilon^{\dagger\nu'}(n)L^k_{\mu'\nu'}
g_{mn'}g_{nn'}\epsilon^{\dagger\mu}(m')\epsilon^\nu(n')H^{ij}_{\mu\nu}
\label{deriv}\\ 
&=&L^k(m,n)g_{mm'}g_{nn'}
\left(
\epsilon^{\dagger\mu}(m')\epsilon_2^{\dagger\alpha}(r)T^i_{\mu\alpha}
\right)
\left(
\epsilon^{\dagger\nu}(n')\epsilon_2^{\dagger\alpha}(s)T^j_{\nu\beta}
\right)^\dagger\nonumber\\
&\times& p_3\epsilon_2(r)\cdot  p_3\epsilon_2^\dagger(s)
\frac{3\,{\rm Br}(D^\ast\to D\pi)}{2\,|{\bf p_3}|}
\nonumber\\
&=&\frac{3\,{\rm Br}(D^\ast\to D\pi)}{2\,|{\bf p_3}|} \, 
\biggl( L^k(t,t)|H^{ij}(t)|^2\cdot (p_3\epsilon_2^\dagger(0))^2
\nonumber\\
&+&\sum\limits_{m,n=\pm,0}L^k(m,n)H^i(m)H^{\dagger j}(n)
\cdot p_3\epsilon_2(m)\cdot p_3\epsilon_2^\dagger(n)
\nonumber\\
&-&\sum\limits_{n=\pm,0}L^k(t,n)H^i(t)H^{\dagger j}(n)
\cdot p_3\epsilon_2(0)\cdot p_3\epsilon_2^\dagger(n)\nonumber\\
&-&\sum\limits_{m=\pm,0}L^k(m,t) H^i(m)H^{\dagger j}(t)
\cdot p_3\epsilon_2(m)\cdot p_3\epsilon_2^\dagger(0)
\biggr)\nonumber
\end{eqnarray}

Using these results one obtains  the full four-fold angular
decay distribution 
\begin{eqnarray}
&&\frac{d\Gamma(B_c\to D^\ast(\to D\pi)\bar l l)}
     {dq^2\,d\cos\theta\,d(\chi/2\pi)\,d\cos\theta^\ast} =
{\rm Br}(D^\ast\to D\pi)
\label{distr4}\\ 
\nonumber\\ 
&&\times
\left\{
\frac{3}{8}\,(1+\cos^2\theta)\cdot\frac{3}{4}\,\sin^2\theta^\ast\cdot
\frac{1}{2}\left(\frac{d\Gamma_U^{11}}{d q^2}+
                   \frac{d\Gamma_U^{22}}{d q^2}\right)
\right.
\nonumber\\
&&
+\frac{3}{4}\,\sin^2\theta\cdot\,\frac{3}{2}\,\cos^2\theta^\ast\cdot
\frac{1}{2}\left(\frac{d\Gamma_L^{11}}{d q^2}+
                   \frac{d\Gamma_L^{22}}{d q^2}\right)\nonumber\\
&&
-\frac{3}{4}\,\sin^2\theta\cdot \cos 2\chi\cdot
 \frac{3}{4}\,\sin^2\theta^\ast\cdot
 \frac{1}{2}\left(\frac{d\Gamma_T^{11}}{d q^2}+
                  \frac{d\Gamma_T^{22}}{d q^2}\right)
\nonumber\\
&&
+\frac{9}{16}\,\sin 2\theta\cdot \cos\chi\cdot\sin 2\theta^\ast\cdot
 \frac{1}{2}\left(\frac{d\Gamma_I^{11}}{d q^2}+
                  \frac{d\Gamma_I^{22}}{d q^2}\right)
\nonumber\\
&&
+v\,
\left[-\frac{3}{4}\,\cos\theta\cdot\,\frac{3}{4}\,
\sin^2\theta^\ast\cdot\frac{d\Gamma_P^{12}}{d q^2}
\right.
\nonumber\\
&&
-\frac{9}{8}\,\sin\theta\cdot \cos\chi\cdot\sin 2\theta^\ast\cdot
 \frac{1}{2}\left(\frac{d\Gamma_A^{12}}{d q^2}+
                    \frac{d\Gamma_A^{21}}{d q^2}\right)
\nonumber\\
&&
\left.
+\frac{9}{16}\,\sin \theta\cdot\sin\chi\cdot\sin 2\theta^\ast\cdot
\left(\frac{d\Gamma_{II}^{12}}{d q^2}
+\frac{d\Gamma_{II}^{21}}{d q^2}\right)\right]\nonumber\\
&&
-\frac{9}{32}\,\sin 2\theta\cdot\sin\chi\cdot\sin 2\theta^\ast\cdot 
\left(\frac{d\Gamma_{IA}^{11}}{d q^2}+\frac{d\Gamma_{IA}^{22}}{d q^2}
\right)\nonumber\\
&&
+\frac{9}{32}\,\sin^2\theta\cdot\sin 2\chi\cdot\sin^2\theta^\ast\cdot 
\left(\frac{d\Gamma_{IT}^{11}}{d q^2}+\frac{d\Gamma_{IT}^{22}}{d q^2}
\right)\nonumber\\
&&
+\frac{3}{4}\,\sin^2\theta\cdot\frac{3}{4}\,\sin^2\theta^\ast\cdot
 \frac{1}{2}\cdot\frac{d\tilde\Gamma_U^{11}}{d q^2}
 - \frac{3}{8}\,(1+\cos^2\theta)\cdot\frac{3}{4}\,\sin^2\theta^\ast\cdot
   \frac{d\tilde\Gamma_U^{22}}{d q^2}
\nonumber\\
&&
+\frac{3}{2}\,\cos^2\theta\cdot\frac{3}{2}\,\cos^2\theta^\ast\cdot
 \frac{1}{2}\cdot\frac{d\tilde\Gamma_L^{11}}{d q^2}
-\frac{3}{4}\,\sin^2\theta\cdot\frac{3}{2}\,\cos^2\theta^\ast\cdot
    \frac{d\tilde\Gamma_L^{22}}{d q^2}
\nonumber\\
&&
+\frac{3}{4}\,\sin^2\theta\cdot \cos 2\chi\cdot
 \frac{3}{4}\,\sin^2\theta^\ast\cdot
 \left(\frac{d\tilde\Gamma_T^{11}}{d q^2}+
       \frac{d\tilde\Gamma_T^{22}}{d q^2}\right)
\nonumber\\
&&
-\frac{9}{8}\,\sin 2\theta\cdot \cos\chi\cdot \sin 2\theta^\ast\cdot
  \frac{1}{2}\left(\frac{d\tilde\Gamma_I^{11}}{d q^2}+
                     \frac{d\tilde\Gamma_I^{22}}{d q^2}\right)
+\frac{3}{2}\,\cos^2\theta^\ast\cdot \frac{1}{4}
 \frac{d\tilde\Gamma_S^{22}}{d q^2}
\nonumber\\
&&
+\frac{9}{16}\,\sin 2\theta\cdot\sin\chi\cdot\sin 2\theta^\ast\cdot 
\left(\frac{d\Gamma_{IA}^{11}}{d q^2}
+\frac{d\Gamma_{IA}^{22}}{d q^2}\right)
\nonumber\\
&&
\left.
-\frac{9}{16}\,\sin^2\theta\cdot\sin 2\chi\cdot\sin^2\theta^\ast\cdot 
\left(\frac{d\Gamma_{IT}^{11}}{d q^2}
+\frac{d\Gamma_{IT}^{22}}{d q^2}\right)
\nonumber
\right\}
\end{eqnarray} 
Integrating Eq.~(\ref{distr4}) over $\cos\theta^\ast$ and 
$\chi$ one recovers the two-fold ($q^2,\cos\theta$) distribution of
Eq.~(\ref{distr2}). 
Note that a similar four-fold distribution has also been obtained in 
Refs.(\cite{Kruger:1999xa},\cite{Kim:2000dq},\cite{Ali:2002qc},
\cite{Chen:2002},\cite{Melikhov:1998cd}) using, however, the zero lepton 
mass approximation. If there are sufficient data one can attempt to fit 
them to the full four-fold decay distribution and thereby extract the 
values of the coefficient functions $d\Gamma_X/dq^2$ and, in the case 
$l=\tau$ the coefficient functions $d\tilde\Gamma_X/dq^2$. 
Instead of considering the full four-fold decay distribution 
one can analyze single angle distributions by integrating out 
two of the remaining angles as done in Ref.~\cite{Faessler:2002ut}.

\section{Model form factors}
\label{sec:mod}

We will employ the relativistic constituent quark model 
\cite{Ivanov:1996pz,Ivanov:1999ic} to calculate the form factors 
relevant to the decays $B\to K \bar ll$ and $B_c\to D(D^\ast) \bar ll$. 
This model is based on an effective interaction Lagrangian which 
describes the coupling between hadrons and their constituent quarks.

For example, the coupling of the meson $H$ to its constituent quarks  
$q_1$ and $\bar q_2 $ is given by the Lagrangian 
\begin{equation}
\label{lag}
{\cal L}_{{\rm int}} (x)=g_H H(x) \int\!\! dx_1 \!\!\int\!\! dx_2
F_H (x,x_1,x_2) \bar q(x_1) \Gamma_H \lambda_H q(x_2)\,.
\end{equation}
Here, $\lambda_H$ and $\Gamma_H$ are  Gell-Mann and Dirac
matrices  which entail the flavor and spin quantum numbers
of the meson field $H(x)$. The function $F_H$ is related to the scalar
part of the Bethe-Salpeter amplitude and characterizes the finite size
of the meson. The function $F_H$ must be invariant under 
the translation  $ F_H(x+a,x_1+a,x_2+a)=F_H(x,x_1,x_2) $.

In our previous papers we have used the so-called impulse approximation 
for the evaluation of the Feynman diagrams. In the impulse approximation
one omits a possible dependence of the vertex functions on external 
momenta. The impulse approximation therefore entails a certain dependence
on how loop momenta are routed through the diagram at hand. This problem 
no longer exists in the present full treatment where the impulse 
approximation is no longer used. In the present calculation we will use 
a particular form of the vertex function given by 
\begin{equation}
\label{vertex}
F_H(x,x_1,x_2)=
\delta\biggl(x-\frac{m_1x_1+m_2x_2}{m_1+m_2}\biggr) \Phi_H((x_1-x_2)^2).
\end{equation}
where $m_1$ and $m_2$ are the constituent quark masses. The vertex 
function $F_H$ evidently satisfies the above translational invariance 
condition. As mentioned before we no longer use the impulse 
approximation in the present calculation.

The coupling constants $g_H$ in Eq.~(\ref{lag}) are determined  
by the so called {\it compositeness condition} proposed in \cite{SWH} 
and extensively used in \cite{Efimov:zg}. The compositeness condition means 
that the renormalization constant of the meson field is set equal to zero
\begin{equation}
\label{z=0}
Z_H=1-\frac{3g^2_H}{4\pi^2}\tilde\Pi^\prime_H(m^2_H)=0
\end{equation}
where $\tilde\Pi^\prime_H$ is the derivative of the meson mass operator.
For the pseudoscalar and vector mesons treated in this paper one has

\begin{eqnarray*}
\tilde\Pi'_P(p^2)&=& \frac{1}{2p^2}\,p^\alpha\frac{d}{dp^\alpha}\,
\int\!\! \frac{d^4k}{4\pi^2i} \tilde\Phi^2_P(-k^2)\\ 
&\times&{\rm tr} \biggl[\gamma^5 S_1(\not\! k+w_{21} \not\!p) \gamma^5 
                         S_2(\not\! k-w_{12} \not\!p) \biggr]\\
&&\\
\tilde\Pi'_V(p^2)&=&
\frac{1}{3}\biggl[g^{\mu\nu}-\frac{p^\mu p^\nu}{p^2}\biggr] 
\frac{1}{2p^2}\,p^\alpha\frac{d}{dp^\alpha}\,
\int\!\! \frac{d^4k}{4\pi^2i} \tilde\Phi^2_V(-k^2)\\
&\times&
{\rm tr} \biggl[\gamma^\nu S_1(\not\! k+w_{21} \not\!p) \gamma^\mu 
                           S_2(\not\! k-w_{12}\not\! p)\biggr]
\end{eqnarray*}
where $w_{ij}=m_j/(m_i+m_j)$.

The leptonic decay constant $f_P$ is calculated from 
\begin{equation}
\label{leptonic}
\frac{3g_P}{4\pi^2} \,\int\!\! \frac{d^4k}{4\pi^2i} 
\tilde\Phi_P(-k^2) 
{\rm tr} \biggl[O^\mu S_1(\not\! k+w_{21} \not\!p) \gamma^5 
                        S_2(\not\! k-w_{12} \not\!p) \biggr]
=f_P\,p^\mu.
\end{equation}
The transition form factors  $P(p_1)\to P(p_2),V(p_2)$ 
 can be calculated from the Feynman integral
corresponding to the diagram of Fig.~\ref{fig:bkformf}:

\begin{eqnarray}
\Lambda^{\Gamma^\mu}(p_1,p_2)&=&
\frac{3g_P g_{P'(V)}}{4\pi^2} \,\int\!\! \frac{d^4k}{4\pi^2i}
\tilde\Phi_P(-(k+w_{13}\,p_1)^2)\,
\tilde\Phi_{P'(V)}(-(k+w_{23}\,p_2)^2)\nonumber\\
&\times&
{\rm tr} \biggl[S_2(\not\! k+\not\! p_2) \Gamma^\mu 
 S_1(\not\! k+\not\! p_1)\gamma^5 S_3(\not\! k)\Gamma_{\rm out}\biggr]
\label{triangle}
\end{eqnarray}
where 
$\Gamma^\mu=\gamma^\mu,\,\gamma^\mu\gamma^5,\,$
$i\sigma^{\mu\nu}q_\nu,\,$ or
$i\sigma^{\mu\nu}q_\nu\gamma^5 $ and 
$\Gamma_{\,P',V}=\gamma^5,\,\gamma_\nu\epsilon_2^\nu. $

We use the local quark propagators

\begin{equation}
S_i(\not\! k)=\frac{1}{m_i-\not\! k} \,,
\end{equation}
where $m_i$ is  the constituent quark mass. 
We do not introduce a new notation for constituent quark masses
in order to distinguish  them from the current quark masses used 
in the effective Hamiltonian and Wilson coefficients as described 
in Sec. II because it should always be clear from the context which
set of masses is being referred to.
As discussed in \cite{Ivanov:1996pz,Ivanov:1999ic}, we assume that 

\begin{equation}
\label{conf}
m_H<m_{1}+m_{2}
\end{equation}
in order to avoid the appearance of imaginary parts in  the physical
amplitudes. 

The fit values for the constituent quark masses are taken from our
papers \cite{Ivanov:1996pz,Ivanov:1999ic} and are given in 
Eq.~(\ref{fitmas}).

\begin{equation}
\begin{array}{ccccc}
     m_u        &      m_s        &      m_c       &     m_b &   \\
\hline
$\ \ 0.235\ \ $ & $\ \ 0.333\ \ $ & $\ \ 1.67\ \ $ & $\ \ 5.06\ \ $ & 
$\ \ {\rm GeV} $\\
\end{array}
\label{fitmas}
\end{equation}
It is readily seen that the constraint Eq.~(\ref{conf}) holds true for 
the low-lying flavored pseudoscalar mesons but is no longer true
for the vector mesons.  
In the case of the heavy mesons $D^\ast$ and $B^\ast$ we
will employ identical  masses for the vector mesons
and the pseudoscalar mesons for the  calculation of
matrix elements in Eqs.~(\ref{z=0}),(\ref{leptonic}) and 
(\ref{triangle}).
It is a quite reliable approximation  because of
$(m_{D^\ast}-m_D)/m_D\sim 7\%$ and 
$(m_{B^\ast}-m_B)/m_B\sim 1\%$.  
In this vein, 
our model was successfully developed for the study of light 
hadrons (e.g., pion, kaon, baryon octet, $\Delta$-resonance), 
heavy-light hadrons (e.g., $D$, $D_s$, $B$ and $B_s$-mesons, 
$\Lambda_Q$, $\Sigma_Q$, $\Xi_Q$ and $\Omega_Q$-baryons) 
and double heavy hadrons (e.g, $J/\Psi$, $\Upsilon$ and 
$B_c$-mesons, $\Xi_{QQ}$ and $\Omega_{QQ}$ 
baryons) \cite{Ivanov:1996pz,Ivanov:1999ic}.  
To extend our approach to other hadrons we had to introduce 
extra model parameters or do some approximations, like, e.g., 
to introduce the cutoff parameter for external hadron momenta 
to guarantee the fulfillment of the mentioned above 
"threshold inequality". Therefore, at the present stage 
we can not apply our approach for the study of rare decays 
involving $K^\ast$ mesons. Probably, it will be a subject 
of our future investigations.

We employ a Gaussian for the vertex function 
$\tilde\Phi_H(k^2_E/\Lambda^2_H) = \exp(-k^2_E/\Lambda^2_H)$ 
where $k_E$ is the Euclidean momentum and 
determine the size parameters $\Lambda^2_H$ by a fit to the 
experimental data, when available, or to lattice simulations for 
the leptonic decay constants. The quality of the fit can be seen 
from Table~\ref{tab:fit}. The branching ratios of the semileptonic 
decays are shown in Table~\ref{tab:sem}. The numerical values 
for $\Lambda_H$ are $\Lambda_\pi=1$~GeV, $\Lambda_K=1.6$~GeV,
$\Lambda_D=2$~GeV and $\Lambda_B=2.25$~GeV for all $K$, $D$ and 
$B$ partners, respectively. 

\begin{table}[ht]
\caption{
Leptonic decay constants $f_H$ (MeV) used in the least-square fit.
The values are taken either from PDG \cite{Hagiwara:in} or
from the Lattice \cite{Ryan}: quenched (upper line)
\newline
and  unquenched (lower line). }
\begin{center}
\def\arraystretch{1.5}
\begin{tabular}{|l|l|l|}
\hline
Meson & This model & Expt/Lattice \\
\hline
$\pi^+$      & 131        & $130.7\pm 0.1\pm 0.36$    \\
\hline
$K^+$        & 161        & $159.8\pm 1.4\pm 0.44$        \\
\hline
$D^+$        & 211        & 203$\pm$ 14 \\
             &            & 226$\pm$ 15 \\
\hline 
$D^+_s$      & 222        &  230$\pm$ 14 \\
             &            &  250$\pm$ 30 \\
\hline   
$B^+$        & 180        &  173$\pm$ 23 \\
             &            &  198$\pm$ 30 \\
\hline   
$B^0_s$      & 196        &  200$\pm$ 20 \\
             &            &  230$\pm$ 30 \\
\hline
$B^+_c$      & 398        &         \\
\hline
\end{tabular}
\label{tab:fit}
\end{center}
\end{table}

\begin{table}[ht]
\caption{
Semileptonic decay branching ratios.
}
\begin{center}
\def\arraystretch{1.5}
\begin{tabular}{|l|l|l|}
\hline
Meson & This model & Expt. \\
\hline
$\pi^+\to \pi^0 l^+ \nu$         & $1.03\cdot 10^{-8}$         & 
                           $(1.025\pm 0.034)\cdot 10^{-8}$    \\
\hline
$K^+  \to \pi^0 l^+ \nu$         & $4.62\cdot 10^{-2}$         & 
                           $(4.82\pm 0.06)\cdot 10^{-2}$    \\
\hline
$B^+\to \bar D^0 l^+ \nu$         & $2.40\cdot 10^{-2}$         & 
                           $(2.15\pm 0.22)\cdot 10^{-2}$    \\
\hline
$B^+\to \bar D^{\ast\,0} l^+ \nu$ & $5.60\cdot 10^{-2}$         & 
                           $(5.3\pm 0.8)\cdot 10^{-2}$    \\
\hline
$B_c^+\to  D^0 l^+ \nu$         & $2.05\cdot 10^{-5}$         & \\ 
\hline
$B_c^+\to  D^{\ast\,0} l^+ \nu$ & $3.60 \cdot 10^{-5}$        & \\
\hline
\end{tabular}
\label{tab:sem}
\end{center}
\end{table}

We are now in a position to present our results for the
$B(B_c)\to K(D,D^\ast)$ form factors. 
We have used the technique outlined 
in our previous papers \cite{Ivanov:1996pz,Ivanov:1999ic}
for the numerical  evaluation  of the Feynman integrals 
in Eq.~(\ref{triangle}).
The results of our numerical calculations are well represented
by the parametrization
\begin{equation}
F(s)=\frac{F(0)}{1-a s+b s^2}\,.
\end{equation} 
Using such a  parametrization facilitates further integrations.
The values of $F(0)$, $a$ and $b$ are listed  in 
Tables~\ref{tab:ff}. 

\begin{table}[ht]
\caption{Parameter values for the approximated form factors \newline 
$F(s)=F(0)/(1-a s + b s^2)$\, $(s=q^2/m_B^2)$.}
\begin{center}
\def\arraystretch{1.5}
\begin{tabular}{|c|ccc|}
\hline
\, $B\to K \bar l l$ \,& $F_+$ & $F_-$ & $F_T$ \\ 
\hline
$F(0)$ & 0.357  & -0.275 & 0.337 \\
$a$    & 1.011  &  1.050 & 1.031 \\
$b$    & 0.042  &  0.067 & 0.051 \\
\hline
\end{tabular}
\end{center}

\begin{center}
\def\arraystretch{1.5}
\begin{tabular}{|c|ccc|cccc|ccc|}
\hline
\, $B_c\to D(D^\ast) \bar l l$ \, & 
$F_+$ & $F_-$ & $F_T$ & $A_0$ & $A_+$ & $A_-$ & $V$ & 
   $a_0$ & $a_+$ & $g$ \\ 
\hline
$F(0)$ & 0.186  & -0.190 & 0.275 & 0.279  & 0.156 & 
        -0.321  &  0.290 & 0.178 & 0.178  & 0.179 \\
$a$    & 2.48   &  2.44  & 2.40  & 1.30   & 2.16  & 
         2.41   &  2.40  & 1.21  & 2.14   & 2.51   \\
$b$    & 1.62   &  1.54  & 1.49  & 0.149  & 1.15  & 
         1.51   &  1.49  & 0.125 & 1.14   & 1.67   \\
\hline
\end{tabular}
\label{tab:ff}
\end{center}
\end{table}

At the end of this section we would like to discuss 
the impulse approximation used in our previous papers
\cite{Ivanov:1996pz,Ivanov:1999ic}. It was simply assumed
that the vertex functions depend only on the loop momentum
flowing through the vertex. The explicit translational
invariant vertex function in Eq.~(\ref{vertex}) allows one
to check the reliability of this approximation. 
We found that the results obtained with and without the impulse
approximation are rather close to each other except for the
heavy-to-light form factors. We consider 
the $B\to\pi$-transition as an example to illustrate this point.
The calculated values of the $F_+^{B\pi}(q^2)$ form factor
at $q^2=0$ are 
\[
F_+^{B\pi}(0)=\left\{  
\begin{array}{ll} 0.27 & \mbox{exact} \\
& \\
 0.48 & \mbox{impulse approximation}
\end{array}
\right.
\]
One can see that the value of the form factor at  $q^2=0$ calculated 
without the impulse approximation is considerably smaller than when 
calculated with the impulse approximation.
Its value is  close to the value of QCD SR estimates,
see, for example, \cite{Bagan:1997bp}:  
$F_+^{B\pi}(0)=0.30$.

\section{Numerical results}

We list our numerical results for the branching ratios in 
Table~\ref{tab:branching}. 
When comparing the values of the branching ratios with 
those obtained in \cite{Ali:1999mm} and \cite{Geng:2001vy}
one finds that they almost agree with each other. 

\begin{table} 
\caption{Decay branching ratios without(with) long distance
contributions.}
\begin{center}
\begin{tabular}{|c|c|c|c|}
\hline
Ref. &
 ${\rm Br}(B\to K\,\mu^+\mu^-)$ & ${\rm Br}(B\to K\,\tau^+\tau^-)$ 
 & ${\rm Br}(B\to K\,\bar\nu\nu)$ \\
\hline\hline
\cite{Ali:1999mm} & $0.57\cdot 10^{-6}$ &$1.3\cdot 10^{-7}$ & \\
\hline
\cite{Ali:2001jg} & 
$(0.35\pm 0.12)\cdot 10^{-6}$ &  & \\ 
\hline
\cite{Melikhov:1997wp} &$0.44\cdot 10^{-6}$ & $1.0\cdot 10^{-7}$ 
& $5.6\cdot 10^{-6}$  \\ 
\hline
\cite{Geng:1996} & $0.5\cdot 10^{-6}$ & $1.3\cdot 10^{-7}$ & \\
\hline
our & $0.55\,(0.51)\cdot 10^{-6}$ & $1.01\,(0.87)\cdot 10^{-7}$ 
& $4.19\cdot 10^{-6}$ \\
\hline
\end{tabular}
\end{center}

\begin{center}
\begin{tabular}{|c|c|c|}
\hline
      & our & \cite{Geng:2001vy} \\
\hline\hline
${\rm Br}(B_c\to D_d\,\mu^+\mu^-)$ &        $0.44\,(0.38)\cdot 10^{-8}$ &  
                                            $0.41\,(0.33)\cdot 10^{-8}$ \\
${\rm Br}(B_c\to D_d^\ast\,\mu^+\mu^-)$ &   $0.71\,(0.58)\cdot 10^{-8}$ &  
                                            $1.01\,(0.78)\cdot 10^{-8}$ \\
${\rm Br}(B_c\to D_s\,\mu^+\mu^-)$ &        $0.97\,(0.86)\cdot 10^{-7}$ &  
                                            $1.36\,(1.12)\cdot 10^{-7}$ \\
${\rm Br}(B_c\to D_s^\ast\,\mu^+\mu^-)$ &   $1.76\,(1.41)\cdot 10^{-7}$ &  
                                            $4.09\,(3.14)\cdot 10^{-7}$ \\
\hline
${\rm Br}(B_c\to D_d\,\tau^+\tau^-)$ &      $0.11\,(0.09)\cdot 10^{-8}$ &  
                                            $0.13\,(0.11)\cdot 10^{-8}$ \\
${\rm Br}(B_c\to D_d^\ast\,\tau^+\tau^-)$ & $0.11\,(0.08)\cdot 10^{-8}$ &  
                                            $0.18\,(0.13)\cdot 10^{-8}$ \\
${\rm Br}(B_c\to D_s\,\tau^+\tau^-)$ &      $0.22\,(0.18)\cdot 10^{-7}$ &  
                                            $0.34\,(0.27)\cdot 10^{-7}$ \\
${\rm Br}(B_c\to D_s^\ast\,\tau^+\tau^-)$ & $0.22\,(0.15)\cdot 10^{-7}$ &  
                                            $0.51\,(0.34)\cdot 10^{-7}$ \\
\hline\hline
${\rm Br}(B_c\to D_d\,\bar\nu\nu)$       & $3.28\cdot 10^{-8}$ &\\
${\rm Br}(B_c\to D_d^\ast\,\bar\nu\nu)$  & $5.78\cdot 10^{-8}$ &\\
${\rm Br}(B_c\to D_s\,\bar\nu\nu)$       & $0.73\cdot 10^{-6}$ &\\
${\rm Br}(B_c\to D_s^\ast\,\bar\nu\nu)$  & $1.42\cdot 10^{-6}$ &\\
\hline
\end{tabular}
\label{tab:branching}
\end{center}
\end{table}


\clearpage


\end{document}